\documentclass[pra,twocolumn,amsmath,amsfonts,amssymb,a4paper,floatfix]{revtex4}
\usepackage{amsthm}
\usepackage{graphicx}

\newcommand{\bbar}[1]{\Bar{\Bar{#1}}}

\begin{document}

\title{Generalized transformation optics from triple spacetime metamaterials}
\author{Luzi Bergamin}
\email{Luzi.Bergamin@esa.int}
\affiliation{European Space Agency, The Advanced Concepts Team (DG-PI), Keplerlaan 1, 2201 AZ Noordijk, The Netherlands}
\date{October 27, 2008}

\begin{abstract}
In this paper, various extensions of the design strategy for transformation media are proposed. We show that it is possible to assign different transformed spaces to the field strength tensor (electric field and magnetic induction) and to the excitation tensor (displacement field and magnetic field), resp. In this way, several limitations of standard transformation media can be overcome. In particular, it is possible to provide a geometric interpretation of non-reciprocal as well as indefinite materials. We show that these transformations can be complemented by a continuous version of electric-magnetic duality and comment on the relation to the complementary approach of field-transforming metamaterials.
\end{abstract}


\maketitle

\section{Introduction}
In the field of metamaterials, artificial electromagnetic materials, the use of spacetime transformations as a design tool for new materials has been proved very successful recently \cite{Pendry:2006Sc,Leonhardt:2006Sc,Leonhardt:2006Nj}. As basic idea of this concept a metamaterial mimics a transformed, but empty space. The light rays follow the trajectories according to Fermat's principle in this transformed (electromagnetic) space instead of laboratory space. This allows one to design in an efficient way materials with various characteristics such as invisibility cloaks \cite{Pendry:2006Sc,Leonhardt:2006Sc,Schurig:2006Sc}, perfect lenses \cite{Leonhardt:2006Nj}, magnification devices \cite{Schurig:2007Oe}, an optical analogue of the Aharonov-Bohm effect or even artificial black holes \cite{Leonhardt:2006Nj}. Still the media relations accessible in this way are rather limited, in particular non-reciprocal or indefinite media (materials exhibiting strong anisotropy) are not covered. But these types of materials also have been linked to some of the mentioned concepts, in particular perfect lenses \cite{Smith:2003Pr,Smith:2004Im} and hyperlenses \cite{Jacob:2006Oe}. This raises the question whether there exists an extension of the concept of transformation media such as to cover those materials as well and to provide a geometric interpretation thereof.

In this paper we propose an extension of this type. As in Refs.\ \cite{Pendry:2006Sc,Leonhardt:2006Sc,Leonhardt:2006Nj} our concept is based on diffeomorphisms locally represented as coordinate transformations. Therefore many of our result allow a geometric interpretation similar to the one of Refs.\ \cite{Leonhardt:2006Nj,Leonhardt:2008Oe} as opposed to another recently suggested route to overcome the restrictions of diffeomorphism transforming media \cite{Tretyakov:2007Me,Tretyakov:2008Gf}. The starting point of our considerations are Maxwell's equations in possibly curved, but vacuous space \footnote{Here and in the following we use Einstein's summation convention, in which a summation over all repeated indices is assumed: $A^i B_i = \sum_i A^i B_i$. For Latin indices this sum runs over the values 1,2,3 (spatial indices), while for Greek indices it runs from 0--4 with $x^0=t$ being time.}:
\begin{align}
\label{maxwell}
 \nabla_i B^i &= 0\ , & \nabla_0 B^i + \epsilon^{ijk}\partial_j E_k &= 0\ , \\
\label{maxwell2}
 \nabla_i \mathcal D^i &= \rho\ , & \epsilon^{ijk} \partial_j \mathcal H_k - \nabla_0 \mathcal D^i &= j^i\ .
\end{align}
Here, $\nabla_i$ is the covariant derivative in three dimensions
\begin{equation}
\label{covder}
 \nabla_i A^i = (\partial_i + \Gamma^i_{i j}) A^j = \frac{1}{\sqrt{\gamma}} \partial_i(\sqrt{\gamma}A^i)\ ,
\end{equation}
with the space metric $\gamma_{ij}$ and its determinant $\gamma$.

For many manipulations it will be advantageous to use relativistically covariant quantities. Therefore, Eqs.\ \eqref{maxwell} and \eqref{maxwell2} are rewritten in terms of the field strength tensor $F_{\mu\nu}$, the excitation tensor $H^{\mu\nu}$ and a four current $J^\mu$ (cf.\ Appendix \ref{sec:conventions}):
\begin{align}
\label{EOMcomp}
 \epsilon^{\mu\nu\rho\sigma} \partial_\nu F_{\rho \sigma} &= 0\ , & D_\nu H^{\mu \nu} &= - J^\mu\ , & D_\mu J^\mu &= 0\ .
\end{align}
The four dimensional covariant derivative $D_\mu$ is defined analogously to \eqref{covder}, whereby the space metric is replaced by the spacetime metric $g_{\mu\nu}$ and its volume element $\sqrt{-g}$.

We wish to analyze these equations of motion from the point of view of transformation media. All transformation materials have in common that they follow as a transformation from a (not necessarily source-free) vacuum solution of the equations of motion, which maps this solution onto a solution of the equations of motion of the transformation material \footnote{Strictly speaking this applies to transformations which are regular everywhere, only. Several singular transformations have been proposed in the literature in the context of metamaterials, e.g.\ the invisibility cloak \cite{Pendry:2006Sc,Schurig:2006Sc,Leonhardt:2006Sc}. In this case a careful study of the global solution is indispensable, as has been done for the case of the cloak in Ref.\ \cite{Greenleaf:2007Cm}.}. The crucial ingredient in the definition of transformation media then is the class of transformations to be considered. As space of all transformations we restrict ourselves to all \emph{linear} transformations in four-dimensional spacetime. Consequently, all media exhibit linear constitutive relations, which may be written within the covariant formulation as \cite{Post}
\begin{equation}
\label{linearH}
 H^{\mu \nu} = \frac{1}{2} \chi^{\mu\nu\rho\sigma} F_{\rho\sigma}\ .
\end{equation}
In vacuum one obtains \footnote{Throughout the paper natural units with $\epsilon_0 = \mu_0 = c = 1$ are used. Notice that the corresponding relation in Ref.\ \cite{Leonhardt:2006Nj} differs from the one used here. According to our conventions $F^{\mu\nu} = H^{\mu\nu}$ in vacuum, while there $\sqrt{-g} F^{\mu\nu} = H^{\mu\nu}$.}
\begin{equation}
\label{chivac}
 \chi^{\mu\nu\rho\sigma} = \frac{1}{2} \left( g^{\mu \rho} g^{\nu \sigma} - g^{\mu \sigma} g^{\nu \rho} \right)\ ,
\end{equation}
such that the standard result $\vec E = \vec{\mathcal D}$ and $\vec B = \vec{\mathcal H}$ emerges.

These transformations and the ensuing media properties \eqref{linearH} have the advantage of being relativistically invariant and thus very easy to handle. However, they do not include any frequency dependence and remain strictly real, which perhaps is the most severe restriction that follows from the coordinate transformation approach. As long as the linear transformations are seen as transformations of spacetime (rather than of the fields) this restriction is not surprising, though. Indeed, from energy conservation it follows that it is impossible to model a process of absorption by the medium as a local transformation of spacetime (notice that the spacetime itself is not dynamical and thus cannot contribute to the energy).

\section{Diffeomorphism transforming metamaterials}
\label{sec:leonhardt}
Obviously, the concept of transformation materials as sketched above is related to symmetry transformations, as those are by definition linear transformations that map a solution of the equations of motion onto another one. Therefore it is worth working out this relation in some more detail.

A symmetry is a transformation which leaves the source-free \footnote{Sources are external parameters and thus should be set to zero for a symmetry transformation. Even together with sources the symmetry can be restored, if an appropriate transformation rule of the sources is defined.} action of the theory, here
\begin{equation}
\label{action}
 \mathcal S =  \int d^4x \sqrt{-g} F_{\mu \nu} H^{\mu \nu}\ ,
\end{equation}
invariant, whereby surface terms are dropped. It straightforwardly follows that a symmetry transformation applied on a solution of the equations of motion still solves the latter. In the above action a general, not necessarily flat, spacetime is considered. The symmetries of this action are well known: these are the $\mbox{\slshape U}(1)$ gauge symmetry of electromagnetism and the symmetries of spacetime (diffeomorphisms). The gauge symmetry cannot help in designing materials as the media relations are formulated exclusively in terms of gauge invariant quantities. However, diffeomorphisms change the media relations, as is pointed out e.g.\ in Ref.\ \cite{Landau2} and as it has been applied to metamaterials in Ref.\ \cite{Leonhardt:2006Nj}. Thus one way to define transformation media is:
\newtheorem{definition}{Definition}

\begin{definition} \label{def:one} A transformation material follows from a symmetry transformation applied to a vacuum solution of Maxwell's equations. This vacuum solution need not be source free. \end{definition}

The space of all possible transformation materials of this kind has been derived in Ref.\ \cite{Leonhardt:2006Nj}; here we briefly want to summarize the result of that paper. The starting point is the observation that a curved space in Maxwell's equations looks like a medium. Indeed, in empty but possibly curved space the constitutive relation among the electromagnetic fields is found by exploiting
\begin{align}
\label{Foi}
 F_{0i} &= (g_{00} g_{ij} - g_{0j}g_{i0})H^{0j} + g_{0k} g_{il} H^{kl}\ , \\
\label{Hij}
 H^{ij} &= 2 g^{i0} g^{jk} F_{0k} + g^{ik} g^{jl} F_{kl}\ ,
\end{align}
which in terms of the space vectors reads
\begin{align}
\label{epsorig}
 \mathcal D^i &= \frac{g^{ij}}{\sqrt{-g_{00}}} E_j - \frac{g_{0j}}{g_{00}} \epsilon^{jil} \mathcal H_l\ , \\
\label{muorig}
 B^i &= \frac{g^{ij}}{\sqrt{-g_{00}}} \mathcal H_j + \frac{g_{0j}}{g_{00}} \epsilon^{jil} E_l\ .
\end{align}
Thus empty space can appear like a medium with permeability and permittivity $\epsilon^{ij} = \mu^{ij} = g^{ij}/\sqrt{-g_{00}}$ and with bi-anisotropic couplings $\xi^{ij} = -\kappa^{ij} = \epsilon^{lij} g_{0l}/g_{00}$.

\begin{figure}[t]
\begin{center}
 \includegraphics[width=\linewidth,bb=0 0 256 43]{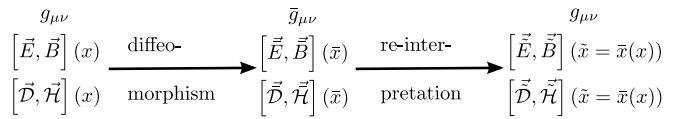}
\end{center}
 \caption{Diffeomorphism transforming metamaterials according to Ref.\ \cite{Leonhardt:2006Nj}.}
 \label{fig:leonhardt}
\end{figure}

Now, as the basic idea of Ref.~\cite{Leonhardt:2006Nj}, if empty space can appear like a medium, a medium should also be able to appear as empty space. One starts with electrodynamics in vacuo, we call these fields $F_{\mu\nu}$ and $H^{\mu\nu}$ with flat metric $g_{\mu\nu}$. Now we apply a diffeomorphism, locally represented as a coordinate transformation $x^{\mu} \rightarrow \bar x^{\mu}(x)$. As the equations of motion by definition are invariant under diffeomorphisms, all relations remain the same with the fields $F$, $H$ and the metric $g_{\mu\nu}$ replaced by the new barred quantities. As a last step one re-interprets in the dynamical equations \eqref{maxwell} and \eqref{maxwell2} the coordinates $\bar x^\mu$ as the original ones $x^\mu$, while keeping $\bar g_{\mu\nu}$ in the constitutive relation. To make this possible some fields must be rescaled in order to transform barred covariant derivatives (containing $\bar g$) into unbarred ones (containing $g$.) The situation of diffeomorphism transforming metamaterials is illustrated in Figure \ref{fig:leonhardt}, which also summarizes our notation. As a more technical remark it should be noted that this manipulation is possible as we consider just Maxwell's theory on a curved background rather than Einstein-Maxwell theory (general relativity coupled to electrodynamics.) In the former case the metric is an external parameter and thus this manipulation is possible as long as none of the involved quantities depends explicitly on the metric.

To keep the whole discussion fully covariant the fields are transformed at the level of the field strength and excitation tensor (rather than at the level of space vectors as was done in Ref.\ \cite{Leonhardt:2006Nj}). To transform the covariant derivatives $\bar D_{\mu}$ into the original $D_{\mu}$ we have to apply the rescalings
\begin{equation}
\label{Hleonhardt}
\tilde H^{\mu\nu} = \frac{\sqrt{-\bar g}}{\sqrt{- g}} \bar H^{\mu\nu} \qquad \tilde J^{\mu} = \frac{\sqrt{-\bar g}}{\sqrt{- g}} \bar J^{\mu}\ .
\end{equation}
In addition the transformation $x^{\mu} \rightarrow \bar x^{\mu}(x)$ may not preserve the orientation of the manifold, which technically means that the Levi-Civita tensor changes sign \cite{Leonhardt:2006Nj}. This is corrected by introducing the sign ambiguity
\begin{equation}
\label{Fsign1}
 \tilde F_{\mu\nu} = \pm \bar F_{\mu\nu}
\end{equation}
with the plus sign for orientation preserving, the minus for non-preserving transformations. These new fields again live in the original space with metric $g_{\mu\nu}$, but now the space is filled with a medium with
\begin{equation}
\label{chisign1}
 \tilde \chi^{\mu \nu \rho \sigma} = \pm \frac{1}{2} \frac{\sqrt{-\bar g}}{\sqrt{- g}} \left( \bar g^{\mu \rho} \bar g^{\nu \sigma} - \bar g^{\mu \sigma} \bar g^{\nu \rho} \right)\ ,
\end{equation}
or, in terms of space vectors,
\begin{align}
\label{epstilde2}
 \tilde{\mathcal D}^i &= s \frac{\bar g^{ij}}{\sqrt{-\bar g_{00}}}  \frac{\sqrt{\bar \gamma}}{\sqrt{\gamma}} \tilde E_j - \frac{\bar g_{0j}}{\bar g_{00}} \epsilon^{jil} \tilde{\mathcal H}_l\ , \\
\label{mutilde2}
 \tilde B^i &= s \frac{\bar g^{ij}}{\sqrt{-\bar g_{00}}} \frac{\sqrt{\bar \gamma}}{\sqrt{\gamma}} \tilde{\mathcal H}_j + \frac{\bar g_{0j}}{\bar g_{00}} \epsilon^{jil} \tilde E_l\ ,
\end{align}
with $s = \pm 1$ being the sign in \eqref{Fsign1} and \eqref{chisign1}. As can be seen, the media properties are restricted to reciprocal materials ($\epsilon = \epsilon^T$, $\mu = \mu^T$, $\kappa = \chi^T$), which, in addition, obey $\epsilon = \mu$. 
This result has been obtained in Ref.\ \cite{Leonhardt:2006Nj} in a slightly different way and encompasses the transformations in Refs.\ \cite{Ward1996:Mo,Pendry:2006Sc}. We do not want to go into further details of this approach but refer to the review \cite{Leonhardt:2008Oe}, where its geometric optics interpretation is discussed in detail. Indeed, light travels in transformation media of this type along null geodesics of the electromagnetic space $\bar x^{\mu}$, which allows (with some restrictions to be discussed in Section \ref{sec:SEMtensor}) a simple and intuitive interpretation of the transformation.

\section{Triple spacetime metamaterials}
\label{sec:triplespace}
Despite the variety of applications of diffeomorphism transforming metamaterials some results suggest a search for extensions. Indeed, there exist e.g.\ designs of super- and hyperlenses that make use of indefinite materials (strong anisotropy) \cite{Smith:2003Pr,Smith:2004Im,Jacob:2006Oe}. Though both concepts should be perfectly understandable in terms of transformation media, the specific material relations used in these works do not fall under the class of diffeomorphism transforming metamaterials.

To understand a possible route to generalize the concept of diffeomorphism transforming media we have to consider again their basis, namely symmetry transformation. The concept of symmetries is used to identify different solutions of the equations of motion that effectively describe the same physics. By means of the re-interpretation in the last step of Figure \ref{fig:leonhardt}, such symmetry transformations can be used as a simple tool to derive within a restricted class of constitutive relations new media properties in a geometrically intuitive and completely algebraic way.

Nonetheless, within the concept of metamaterials it is not important that the transformed solution in principle describes the same physics as the original one. Still, one may want to keep the possibility of mapping source free solutions onto other source free solutions in a straightforward way, as only in this way do we have an effective control over passive media and do not risk introducing exotic sources such as magnetic monopoles. Furthermore a geometric interpretation of the transformations is kept, which is advantageous in many applications. To weaken the conditions on transformation materials while keeping the advantages of symmetry transformations we thus propose the following definition:

\begin{definition}
\label{def:two}
 Consider the set of all transformations $T$ which map a source free solution of the equations of motion \eqref{maxwell} and \eqref{maxwell2} onto another source free solution. A transformation material is a material obtained by applying a transformation $T$ onto a (not necessarily source free) vacuum solution.
\end{definition}

There are two types of extensions contained in this definition compared to the previous section:
\begin{enumerate}
 \item There exist transformations that leave the equations of motion invariant, but change the action by a constant and thus are not symmetry transformations. A transformation of this type is the so-called electric-magnetic duality. Its effect will briefly be discussed in Section \ref{sec:emduality}.
\item We do allow for transformations which leave all Maxwell's equations \eqref{maxwell} and \eqref{maxwell2} invariant, but change the media relations \eqref{linearH}. This indeed generalizes the concept in an important way.
\end{enumerate}

To see the origin of the second extension it is important to realize that the equations of motion of electrodynamics separate into two different sets (Eqs.~\eqref{maxwell} and \eqref{maxwell2}, resp.)\ with mutually exclusive field content. This characteristic is not just an effect of our notation, but as has been shown e.g.\ in Refs.\ \cite{Hehl:2003,Hehl:2005hu}, the equations of motion of electrodynamics can be derived from first principles without using explicitly the constitutive relation $H = H(F)$. As the two sets of equations are separately invariant under diffeomorphisms it should be possible to assign \emph{different} transformed spaces to $H = (\vec{\mathcal D}, \vec{\mathcal H})$ and $F = (\vec{E},\vec{B})$. In other words, it must be possible to distort the spaces (or the coordinates) of the field strength tensor and the excitation tensor separately, whereby the resulting transformation material per constructionem satisfies all conditions of the Definition \ref{def:two}. The ensuing constitutive relation as well as the solutions of the equations of motion still follow (almost) as simple as in the case of Ref.\ \cite{Leonhardt:2006Nj}.

\begin{figure}[t]
\begin{center}
 \includegraphics[width=\linewidth,bb=0 0 256 105]{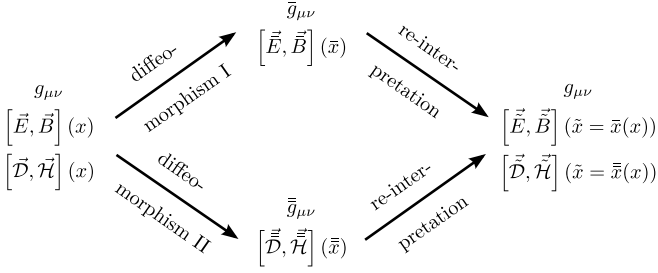}
\end{center}
 \caption{Illustration and notation of the generalized ``triple spacetime metamaterials''. Notice that the diffeomorphism I acts only on the fields $\vec E$ and $\vec B$, while diffeomorphism II acts on $\vec{\mathcal D}$ and $\vec{\mathcal H}$.}
 \label{fig:triplespace}
\end{figure}

To prove the potential of this method we have to extend the notation compared to the previous section: as before laboratory space has metric $g_{\mu\nu}$, its fields in vacuo are $H = (\vec{\mathcal D}, \vec{\mathcal H})$ and $F = (\vec{E},\vec{B})$; the fields of the transformation material (living in the space with metric $g_{\mu\nu}$) are again labeled with a tilde. The transformed space of the field strength tensor has metric $\bar g_{\mu\nu}$ and fields $\bar F = (\vec{\Bar E},\vec{\Bar B})$, the one of the excitation tensor $\bbar{g}_{\mu\nu}$ and $\bbar H = (\vec{\bbar{\mathcal D}}, \vec{\bbar{\mathcal H}})$. This new transformation is illustrated in Figure \ref{fig:triplespace}. Applying the two transformations
\begin{equation}
 \bar x^\mu = \bar x^\mu(x)\ , \qquad \bbar x^\mu = \bbar x^\mu(x)
\end{equation}
to the constitutive relation \eqref{linearH} with $\chi$ being the vacuum relation \eqref{chivac} yields
\begin{equation}
 \bbar{H}^{\mu\nu} = \frac{1}{2} \frac{\partial \bbar{x}^{\mu}}{\partial x^{\lambda}} \frac{\partial \bbar{x}^{\nu}}{\partial x^{\tau}} \left( g^{\lambda \alpha} g^{\tau \beta} - g^{\lambda \beta} g^{\tau \alpha} \right) \frac{\partial \bar{x}^{\rho}}{\partial x^{\alpha}} \frac{\partial \bar{x}^{\sigma}}{\partial x^{\beta}} \bar F_{\rho\sigma}\ .
\end{equation}
Introducing the notation
\begin{equation}
\label{gbbb}
 g^{\bbar{\mu} \bar \nu} = \frac{\partial \bbar{x}^{\mu}}{\partial x^{\rho}} \frac{\partial \bar{x}^{\nu}}{\partial x^{\sigma}} g^{\rho \sigma} = \bbar{g}^{\mu \rho} \frac{\partial \bar{x}^{\nu}}{\partial \bbar x^{\rho}} = \frac{\partial \bbar{x}^{\mu}}{\partial \bar x^{\rho}} \bar g^{\rho \nu}
\end{equation}
the relation may be written as
\begin{equation}
\label{tripelconst}
  \bbar{H}^{\mu\nu} = \frac{1}{2}  \left( g^{\bbar \mu \bar \rho} g^{\bbar \nu \bar \sigma} - g^{\bbar \mu \bar \sigma} g^{\bbar \nu \bar \rho} \right) \bar F_{\rho\sigma}\ .
\end{equation}
It should be noted that $g^{\bbar{\mu} \bar \nu}$ in Eq.\ \eqref{gbbb} is no longer a metric, in particular it need not be symmetric in its indices and it need not have signature $(3,1)$.

To derive the new constitutive relations in the original (laboratory) space we proceed analogously to the previous section. All fields have to be rescaled in order to obey the equations of motion in the original space with metric $g_{\mu\nu}$, which implies
\begin{align}
\label{trtildedef}
 \tilde F_{\mu\nu} &= \pm \bar F_{\mu\nu}\ , & \tilde H^{\mu\nu} &= \frac{\sqrt{-\bbar g}}{\sqrt{- g}} \bbar H^{\mu\nu}\ , & \tilde J^{\mu} &= \frac{\sqrt{-\bbar g}}{\sqrt{-g}} \bbar J^{\mu}\ .
\end{align}
Thus the constitutive relation becomes
\begin{equation}
 \label{xchi}
 \tilde H^{\mu\nu} = \tilde \chi^{\mu\nu\rho\sigma} \tilde F_{\rho\sigma} = \pm \frac12 \frac{\sqrt{-\bbar g}}{\sqrt{- g}} \left( g^{\bbar \mu \bar \rho} g^{\bbar \nu \bar \sigma} - g^{\bbar \mu \bar \sigma} g^{\bbar \nu \bar \rho} \right) \tilde F_{\rho\sigma}\ ,
\end{equation}
where the sign refers to the possible change of orientation in the transformation $x^\mu \rightarrow \bar x^\mu$.
For the equivalent relation in terms of space vectors the notation
\begin{equation}
 \epsilon_{\mu\nu\rho\sigma} = \bar s \frac{\sqrt{-g}}{\sqrt{-\bar g}} \bar \epsilon_{\mu\nu\rho\sigma} =  \bbar s \frac{\sqrt{-g}}{\sqrt{-\bbar g}} \bbar \epsilon_{\mu\nu\rho\sigma}
\end{equation}
is used, where $\bar s$ and $\bbar{s}$ are the respective signs due to the change of orientation in the transformations to laboratory space. Now it easily follows from \eqref{ABdef}--\eqref{CDdef} that
\begin{gather}
\label{tripleA}
 \mathfrak A^{ij} = -\bar s \frac{\sqrt{-\bbar g}}{\sqrt{\gamma}} (g^{\bbar 0 \bar 0}g^{\bbar i \bar j} - g^{\bbar 0 \bar j}g^{\bbar i \bar 0})\ , \\
\label{tripleB}
 \mathfrak B_{ij} = - \bbar s \frac{\sqrt{\gamma}}{\sqrt{-\bar g}} (g_{\bbar 0 \bar 0}g_{\bbar i \bar j} - g_{\bbar 0 \bar j}g_{\bbar i \bar 0})\ , \\
\label{tripleC}
 \mathfrak C_i{}^j = - \frac{\bar s}{2} \frac{\sqrt{-\bbar g}}{\sqrt{\gamma}} \epsilon_{ikl} (g^{\bbar k \bar 0} g^{\bbar l \bar j} - g^{\bbar k \bar j} g^{\bbar l \bar 0})\ , \\
\label{tripleD}
 \mathfrak D^i{}_j =  \frac{\bar s}{2} \frac{\sqrt{-\bbar g}}{\sqrt{\gamma}} \epsilon_{jkl} (g^{\bbar 0 \bar k} g^{\bbar i \bar l} - g^{\bbar 0 \bar l} g^{\bbar i \bar k})\ ,
\end{gather}
which are the defining tensors of the Boys-Post relation. After some algebra the Tellegen relation
\begin{align}
\label{tripeleps}
 \tilde{\mathcal D}^i &= - \bar s \frac{\sqrt{-\bbar g}}{\sqrt{\gamma} g_{\bar 0 \bbar{0}}} g^{\bbar{i} \bar j} \tilde E_j - \bar s \bbar{s} \frac{\sqrt{-\bar g}\sqrt{-\bbar{g}}}{\gamma g_{\bar 0 \bbar{0}}} g^{\bbar{i} \bar k} g^{\bbar{0} \bar l} \epsilon_{klm} g^{\bar m\bbar j} \tilde{\mathcal H}_j\ , \\
\label{tripelmu}
 \tilde B^i &=  - \bbar s \frac{\sqrt{-\bar g}}{\sqrt{\gamma} g_{\bar 0 \bbar{0}}} g^{\bar{i} \bbar j} \tilde{\mathcal H}_j + \bar s \bbar{s} \frac{\sqrt{-\bar g}\sqrt{-\bbar{g}}}{\gamma g_{\bar 0 \bbar{0}}}  g^{\bar{i} \bbar k} \epsilon_{klm} g^{\bbar l \bar 0} g^{\bbar m \bar j} \tilde E_j
\end{align}
is found, which in the limit of $\bbar g_{\mu\nu} = \bar g_{\mu\nu}$ is equivalent to Eqs.\ \eqref{epstilde2} and \eqref{mutilde2}. An important comment is in order: due to the different transformations applied to $H^{\mu\nu}$ and $F_{\mu\nu}$, resp., the constitutive relation \eqref{xchi}, or \eqref{tripeleps} and \eqref{tripelmu}, relates fields from \emph{different} spacetime points in the original space, e.g.\ $\tilde E_i\left(\tilde x = \bar x(x)\right)$ refers the field $E_i(x)$ at a different point $x^\mu$ in the original space than $\tilde{\mathcal D}^i \left(\tilde x = \bbar x(x)\right)$ does.

Let us comment on the more technical parts of this result. In Section \ref{sec:leonhardt} we saw that transformation materials derived from symmetry transformations are restricted to reciprocal materials with $\epsilon = \mu$. These restrictions can be overcome partially with the above result:
\begin{itemize}
 \item As $g^{\bbar i \bar j} = (g^{\bar j \bbar i})^T$ it follows that permittivity and permeability are related as
 \begin{equation}
\label{epsmurel}
  \bbar{s} \sqrt{-\bbar{g}} \mu^{ij} = \bar s \sqrt{-\bar g} \epsilon^{ji}\ .
 \end{equation}
 It should not come as a surprise that permittivity and permeability cannot be independent, as by virtue of the definition of the relativistically covariant tensors $F_{\mu\nu}$ and $H_{\mu\nu}$ such transformations cannot act independently on $\vec E$ and $\vec B$ or $\vec{\mathcal D}$ and $\vec{\mathcal H}$, resp. A possible route to relax this restriction is discussed in Section \ref{sec:fieldtransforming}.
 \item Permittivity and permeability need no longer be symmetric. Therefore it is possible to describe non-reciprocal materials, or, in the language of Eq.\ \eqref{chidecomp}, the skewon part need not vanish. This happens if the mapping between the two electromagnetic spaces, $\partial \bar x^\mu / \partial \bbar x^\nu$, is not symmetric in $\mu$ and $\nu$, e.g.\ for a material with mapping $\bar x = x-z$, $\bbar{x} = x+z$.
\item The generalized transformations yield many more possibilities considering the signs of the eigenvalues of permittivity and permeability. Within the method of Ref.\ \cite{Leonhardt:2006Nj}, $\mu$ and $\epsilon$ are essentially determined by the spatial metric of the electromagnetic space (cf.\ Eqs.\ \eqref{epstilde2} and \eqref{mutilde2} and recall the relation $g^{ij} = \gamma^{ij}$.) However, a spatial metric by definition must have three positive eigenvalues, a characteristic that cannot be changed by any diffeomorphism. Thus it follows that in any medium of this type the eigenvalues of $\epsilon$ and $\mu$ are all of the same sign.
\begin{itemize}
 \item Within the generalized setup of ``triple spacetime metamaterials'', however, the signs of the eigenvalues in  $\epsilon$ can be chosen freely, as the metric is multiplied by a transformation matrix,
 \begin{equation}
  g^{\bbar i \bar j} = \bbar g^{i \mu} \frac{\partial \bar x^j}{\partial \bbar x^\mu}\ ,
 \end{equation}
 and no restrictions on the signs of the eigenvalues of the transformation matrix exist. In this way indefinite materials \cite{Smith:2003Pr,Smith:2004Im} can be designed as a result of different space inversions in the two different mappings. As an example the mapping $\bar z = -z$, $\bbar{z} = z$ (with all other directions mapped trivially) yields $\epsilon^{ij} = \mbox{diag}(-1,-1,1)$, $\mu^{ij} = \mbox{diag}(1,1,-1)$.
\item Furthermore the relative sign between the eigenvalues of $\epsilon$ and those of $\mu$ can be chosen as a consequence of the factor $\bbar{s}$ in Eq.\ \eqref{epsmurel}. This change in the relative sign may be interpreted as a partial reversal of time as can be seen in the following list (space maps trivially here and all media are assumed to be homogeneous):
\[
 \begin{array}{|c||c|c|c|c|}
 \hline &&&&\\[-2.2ex]
 & \bar t & \bbar t & \epsilon & \mu  \\ \hline \hline
 I & t & t & 1 & 1 \\
 II & t & -t & -1 & 1  \\
 III & -t & t & 1 & -1  \\
 IV & -t & -t & -1 & -1 \\ \hline
\end{array}
\]

\end{itemize}
We note that all eight classes of materials discussed in Ref.\ \cite{Smith:2003Pr} allow a geometric interpretation within the setup of ``triple spacetime metamaterials.''
\item More complicated than permittivity and permeability are the bi-anisotropic couplings. With the standard assumption of $g^{0i} = 0$ in laboratory space it follows from 
\begin{align}
 \xi^{ij} &= - \bar s \bbar{s} \frac{\sqrt{-\bar g}\sqrt{-\bbar{g}}}{\gamma g_{\bar 0 \bbar{0}}} g^{\bbar{i} \bar k} g^{\bbar{0} \bar l} \epsilon_{klm} g^{\bar m\bbar j}\ , \\ \kappa^{ij} &= \bar s \bbar{s} \frac{\sqrt{-\bar g}\sqrt{-\bbar{g}}}{\gamma g_{\bar 0 \bbar{0}}}  g^{\bar{i} \bbar k} \epsilon_{klm} g^{\bbar l \bar 0} g^{\bbar m \bar j}\ ,
\end{align}
 similarly to Eq.\ \eqref{epstilde2} that all electric-magnetic couplings vanish if the transformation does not mix space and time. In this case the crucial components $g^{\bar{0} \bbar l}$ and $g^{\bbar{0} \bar l}$ may be written as
\begin{align}
 g^{\bar{0} \bbar l} &= \frac{\partial \bar x^0}{\partial x^0} g^{00} \frac{\partial \bbar x^l}{\partial x^0} + \frac{\partial \bar x^0}{\partial x^i} g^{ij} \frac{\partial \bbar x^l}{\partial x^j}\ ,\\
 g^{\bbar{0} \bar l} &= \frac{\partial \bbar x^0}{\partial x^0} g^{00} \frac{\partial \bar x^l}{\partial x^0} + \frac{\partial \bbar x^0}{\partial x^i} g^{ij} \frac{\partial \bar x^l}{\partial x^j}\ .
 \end{align}
 Most importantly it is found from these expressions that one of the two bi-anisotropic couplings may vanish while the other one is non-vanishing, which is impossible within the context of diffeomorphism transforming media. Moreover, in the latter case the bi-anisotropic couplings must be symmetric matrices, which need no longer be the case in the present context.
 \item Finally, the result \eqref{tripeleps} and \eqref{tripelmu} reduces to the relations \eqref{epstilde2}, \eqref{mutilde2} if $g_{\bar \mu \bbar{\nu}}$ is a symmetric matrix of signature $(3,1)$ and, in addition, $\sqrt{-\bar g} = \sqrt{- \bbar g}$. This does not necessarily imply $\bar x^\mu = \bbar x^\mu$ but rather that there exists yet a different space which describes the same media properties in terms of the transformations of Section \eqref{sec:leonhardt}.

\end{itemize}

\subsection{Electric-magnetic duality and rotation}
\label{sec:emduality}

Finally, we should ask whether Eqs.\ \eqref{tripeleps}, \eqref{tripelmu} indeed describe the most general media fulfilling Definition \ref{def:two}. Taken separately, the two sets of equations in \eqref{maxwell} and \eqref{maxwell2} do not exhibit more symmetries than diffeomorphisms. However, there exists the possibility of transformations that mix $F_{\mu\nu}$ and $H^{\mu\nu}$. Indeed a transformation of this type is known as electric-magnetic duality, which has important implications in modern theoretical high-energy physics \cite{Montonen:1977sn}. It represents the fact that under the exchange
\begin{equation}
 F_{\mu\nu} \leftrightarrow \ast H_{\mu \nu}
\end{equation}
or in terms of space vectors
\begin{align}
 B^i &\rightarrow - \mathcal D^i\ , & \mathcal H_i &\rightarrow - E_i\ , \\ E_i &\rightarrow \mathcal H_i\ , & \mathcal D^i &\rightarrow B^i\ ,
\end{align}
the source-free equations of motion do not change (the action changes by an overall sign.) Of course, this duality transformation is problematic when applied to a solution with sources, as it transforms electric charges and currents into magnetic charges and currents and vice versa. In the remainder of this section we thus restrict to source-free solutions or should allow the possibility of artificial magnetic monopoles. Then it can be checked straightforwardly that electric-magnetic duality applied to the result \eqref{tripelconst}, or \eqref{tripeleps} and \eqref{tripelmu}, does not yield media relations not yet covered by diffeomorphisms alone.

\begin{figure}[t]
\begin{center}
 \includegraphics[width=\linewidth,bb=0 0 275 105]{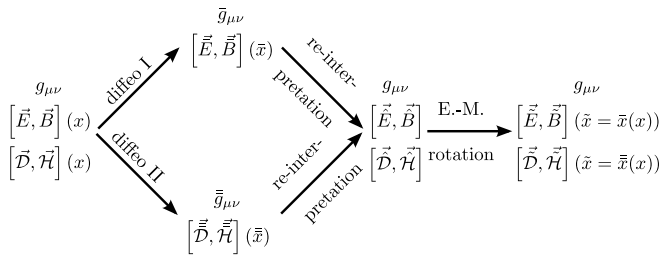}
\end{center}
 \caption{Illustration and notation of the generalized ``triple spacetime metamaterials'' complemented by electric-magnetic rotation. The electric-magnetic rotation must act after the transformation of spacetime as these two steps do not commute.}
 \label{fig:emrotation}
\end{figure}
However, as far as the equations of motion \eqref{maxwell} and \eqref{maxwell2} are concerned, electric-magnetic duality can be promoted to a continuous $U(1)$ symmetry with transformation \footnote{Notice that under the continuous transformation the action behaves as $\mathcal S \rightarrow (\cos^2\alpha - \sin^2\alpha) \mathcal S$ and thus for $\alpha = \pi/4$ transforms to zero. Therefore at the level of the action only the discrete duality transformation can be considered.}
\begin{align}
\label{EMD1}
 \tilde B^i &= \cos \alpha B^i - \sin \alpha \mathcal D^i\ , & \tilde{\mathcal D}^i &= \cos \alpha \mathcal D^i + \sin \alpha B^i\ , \\
\label{EMD2}
 \tilde E_i &= \cos \alpha E_i + \sin \alpha \mathcal H_i\ , & \tilde{\mathcal H}_i &= \cos \alpha \mathcal H_i - \sin \alpha E_i\ .
\end{align}
These transformations comply with Definition \ref{def:two} and thus their action onto a medium with general constitutive relation \eqref{tellegen} should be studied. The result
\begin{align}
 \begin{split}
 \tilde{\mathcal D}^i &= \left(\cos^2 \alpha \epsilon + \sin^2 \alpha \mu + \sin\alpha\cos\alpha (\kappa+\xi) \right)^{ij} \tilde E_j  \\
 &\quad + \left(\cos^2 \alpha \kappa - \sin^2 \alpha \xi + \sin\alpha\cos\alpha (\mu-\epsilon) \right)^{ij} \tilde{\mathcal H}_j\ ,
\end{split}\\
\begin{split}
 \tilde{B}^i &= \left(\cos^2 \alpha \mu + \sin^2 \alpha \epsilon - \sin\alpha\cos\alpha (\kappa+\xi) \right)^{ij} \tilde{\mathcal H}_j  \\
 &\quad + \left(\cos^2 \alpha \xi - \sin^2 \alpha \kappa + \sin\alpha\cos\alpha (\mu-\epsilon) \right)^{ij} \tilde E_j
\end{split}
\end{align}
shows that the transformation acts trivially if $\epsilon = \mu$ and $\xi = -\kappa$, in particular in vacuo and consequently for all diffeomorphism transforming media \eqref{epstilde2}. However, they yield new media relations when acting on a solution of the type \eqref{tripeleps} and \eqref{tripelmu}. Therefore these new relations are part of the materials covered by Definition \ref{def:two}. They are derived here for completeness, though their geometric interpretation is not immediate. The coordinate lines $\bar x^{\mu}(x)$ and $\bbar{x}^{\mu}(x)$ could be understood as the electromagnetic spaces of the linear combinations $(\vec{\tilde E}, \vec{\tilde{B}})$ and $(\vec{\tilde{\mathcal D}}, \vec{\tilde{\mathcal H}})$ as given in \eqref{EMD1} and \eqref{EMD2}, resp. Still, one should be careful with this interpretation: as the transformation of spacetime does not commute with the electric-magnetic rotation one cannot modify the situation in Figure \ref{fig:emrotation} in such a way that the two electromagnetic spaces, $\bar x^\mu$ and $\bbar x^\mu$, are identified with certain linear combinations of $(\vec E,\vec B)$ and $(\vec{\mathcal D}, \vec{\mathcal H})$, resp.; rather the electric-magnetic rotation acts upon the fields after the transformation of spacetime.

\section{Perfect lens from indefinite material: an example}
To provide a better understanding of the formalism developed in the previous section a concrete example is demonstrated. To keep things simple we show how a proposal taken from the literature can be given a geometric interpretation.

In Ref.~\cite{Smith:2003Pr} it has been pointed out that two slabs of indefinite material (media with strong anisotropy) can form a perfect lens. Since, in constrast to standard diffeomorphism transforming media, strong anisotropy is available in triple spacetime metamaterials the question appears whether a geometric interpretation of the lens proposed in Ref.~\cite{Smith:2003Pr} can be given (cf.\ Ref.~\cite{Bergamin:2008Mm} for a related discussion.) We consider the lens to be an infinite slab in the x-y plane with a certain thickness in the z direction. In its simplest form the lens consists of two slabs of equal thickness $d$, where the media properties of the first slab are
\begin{equation}
\label{example1}
 \epsilon^{ij} = \mu^{ij} = \mbox{diag}(1,1,-1)\ ,
\end{equation}
while in the second slab
\begin{equation}
\label{example2}
 \epsilon^{ij} = \mu^{ij} = \mbox{diag}(-1,-1,1)\ .
\end{equation}
To provide a geometric interpretation we start with the observation that a standard perfect lens with $\epsilon=\mu=-1$ may be produced by two different transformations, either a space inversion $\bar z = -z$, or a time reversal $\bar t = -t$. From Eqs.~\eqref{epstilde2} and \eqref{mutilde2} it follows straightforwardly that these two transformations yield the same media properties. Within triple spacetime metamaterials we now may ask the question of what happens if space inversion is applied to one set of the fields, while time reversal is applied to the other set. For concreteness, space inversion is applied to the fields $\vec E$ and $\vec B$ and thus
\begin{equation}
\label{example3}
 \bar z = -z + Z_1\ ,
\end{equation}
where $Z_1$ is an unimportant constant necessary to meet the boundary conditions. All other fields $\bar x^{\mu}$ are mapped trivially. The second set of fields, $\vec{\mathcal D}$ and $\vec{\mathcal H}$, transform according to
\begin{equation}
\label{example4}
 \bbar{t} = -t + T_1
\end{equation}
with all other fields transformed trivially. Consider now these two transformations in Eqs.~\eqref{tripeleps} and \eqref{tripelmu}. From \eqref{gbbb} one finds
\begin{equation}
 g^{\bar i \bbar j} = g^{\bbar i \bar j} = \mbox{diag}(1,1,-1)\ .
\end{equation}
Furthermore, $\bar s = \bbar s = -1$ as both tranformations are orientation changing. Furthermore, $g_{\bar 0 \bbar 0} = 1$ (remember our convention $g_{00} = -1$), such that indeed the media properties \eqref{example1} are found in this slab.

\begin{figure}
 \centering
 \includegraphics[width=\linewidth]{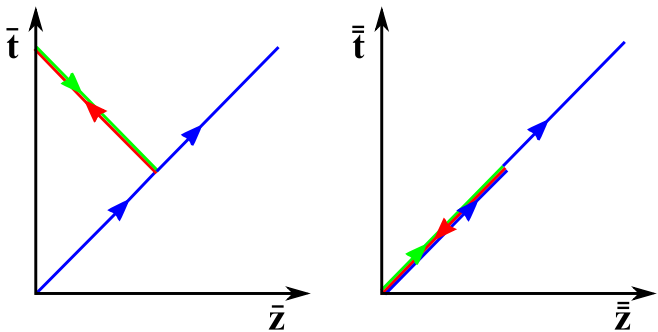}
 \caption{Mapping of the world-line $s(t) = (t,z(t))$ in the original space onto the deformed spaces by means of the two different transformations. Blue lines indicate the trajectory outside of the lens (trivial mapping of all $x^\mu$), the red line represents the first slab, the green line the second slab. The parametrization of the world-line in the original space is assumed to obey $z(t) =  t$.}
 \label{fig:example}
\end{figure}
This single slab of indefinite material does not establish a perfect lens, as can be seen easily when studying how a world-line $s(\tau) = \left(t(\tau),x(\tau),y(\tau),z(\tau)\right)$ is mapped onto the two deformed spaces. For simplicity time may be interpreted with the parametrization variable $t(\tau) = \tau$ and furthermore we can assume without loss of generality $z(\tau) = \tau = t$ \footnote{``Time'' $t$ here is just a variable to parametrize the world line, this choice does not make any statements about the speed of light.}. The situation is illustrated in Figure \ref{fig:example}. As can be seen the mappings do not agree after the first slab, both trajectories are at the same point $z = 0$, but they differ in time. This must be corrected in the second slab. In our example we have chosen a completely trivial mapping for $\vec{\mathcal D}$ and $\vec{\mathcal H}$, so these fields propagate in the second slab as in free space. $\vec E$ and $\vec B$, however, are transformed as
\begin{align}
 \bar{t} &= -t + T_2\ , & \bar{z} &= -z + Z_2\ ,
\end{align}
which actually reverts the transformation \eqref{example3} and at the same time applies \eqref{example4}. Not surprisingly, the two trajectories now meet at the same point again and the perfect lens is established. Again it is immediate that this transformation establishes the media relations \eqref{example2}. Therefore, triple spacetime metamaterials indeed can provide a geometric interpretation of the lens of Ref.~\cite{Smith:2003Pr}. It should be noted, that this specific lens has focal length zero, it shrinks the effective width of the device from $2D$ to zero, but not to a negative value as is necessary for a real lens.

\section{Energy, momentum and wave vector}
\label{sec:SEMtensor}
So far we studied solutions of Maxwell's equations which---up to rescalings---are equivalent to certain vacuum solutions. Still we did not ask up to what point these transformation materials really are ``media that look like empty space.'' To do so it is not sufficient to consider the transformation of the fields and sources, but equally well we should look at the conservation laws, summarized in the conservation of the stress-energy-momentum tensor (SEM tensor.) While in the generic situation of electrodynamics in media, the definition of the ``SEM tensor of electrodynamics'' is not unique \cite{Israel:1980Gr,Dereli:2007Co}, we do not have to deal with these subtleties in the present situation as our (idealized) media are lossless and dispersion free and thus allow for a definition of a complete action (cf.\ Eq.~\eqref{action}) without any reference to ``matter.'' Therefrom we immediately derive the covariant SEM tensor
\begin{gather}
 \label{covSEM}
 T^{\mu\nu} = - \frac{1}{4\pi} \left(F_{\rho\sigma} g^{\sigma\{\mu} H^{\nu\}\rho} + \frac{1}{4} g^{\mu\nu} F_{\rho\sigma}H^{\rho\sigma}\right)\ ,\\ D_\mu T^{\mu\nu} = 0\ .
\end{gather}
The advantage of this tensor over the canonical SEM tensor is the simple behavior under diffeomorphisms: being a real tensor field, $T_{\mu\nu}$ transforms exactly in the same way as the metric.

Let us now look at the materials as described in Section \ref{sec:leonhardt}. Thanks to its transformation properties the SEM tensor in the electromagnetic space follows immediately as
\begin{equation}
 \bar T^{\mu\nu} = - \frac{1}{4\pi} \left(\bar F_{\rho\sigma} \bar g^{\sigma\{\mu} \bar H^{\nu\}\rho} + \frac{1}{4} \bar g^{\mu\nu} \bar F_{\rho\sigma}\bar H^{\rho\sigma}\right)\ .
\end{equation}
But how about $\tilde T_{\mu\nu}$? Of course one could define an ``induced SEM tensor'' from the electromagnetic space as (cf.\ Eqs.\ \eqref{Hleonhardt} and \eqref{Fsign1})
\begin{equation}
 \label{covSEMI}
 \tilde T_I^{\mu\nu} = \mp \frac{\sqrt{- g}}{\sqrt{-\bar g}} \frac{1}{4\pi} \left(\tilde F_{\rho \sigma} \bar g^{\sigma \{\mu}\tilde H^{\nu\}\rho} + \frac{1}{4} \bar g^{\mu\nu} \tilde F_{\rho\sigma}\tilde H^{\rho\sigma}\right)\ ,
\end{equation}
but obviously this tensor is not conserved in laboratory space, $D_\mu \tilde T_I^{\mu\nu} \neq 0$, since it depends explicitly on the metric $\bar g_{\mu\nu}$. In other words, the crucial trick to re-interpret in the dynamical equations the coordinates in electromagnetic space, $\bar x^\mu$, as those in laboratory space, $x^\mu$, works in the equations of motion \eqref{maxwell} and \eqref{maxwell2}, but does not work for the SEM tensor and its conservation.

Of course, the correct SEM tensor in laboratory space immediately follows from \eqref{covSEM} as
\begin{equation}
 \label{covSEMD}
 \tilde T^{\mu\nu} = - \frac{1}{4\pi} \left(\tilde F_{\rho \sigma}  g^{\sigma\{\mu}  \tilde H^{\nu\} \rho} + \frac{1}{4} g^{\mu\nu} \tilde F_{\rho\sigma}\tilde H^{\rho\sigma}\right)\ .
\end{equation}
Clearly, requiring equivalence of the two tensors would not even allow for conformal transformations. But even when looking at integrated quantities (total energy and momentum flux in the material),
\begin{equation}
 P^\mu = \int\! d^3x \sqrt{\gamma}\, T^{0\mu}\ ,
\end{equation}
the induced tensor does not yield the correct quantity in laboratory space. Of course, the situation is even more complicated for triple spacetime metamaterials: since the transformation of the explicit metrics appearing in Eq.\ \eqref{covSEM} is not defined, an ``induced SEM tensor'' cannot even be defined.

Instead of the correct, directly evaluated SEM tensor \eqref{covSEMD} a slightly different tensor is considered in the following. To see its advantage we make the standard assumption that our laboratory space metric has $g_{00} = -1$ ($x^0$ is our laboratory time) and $g_{0i} = 0$ (the measure of distances is time independent.) Then it is straightforward that the quantity
\begin{equation}
\label{mmunu}
 M^{\mu\nu} = - g^{\mu\rho} F_{\rho\sigma} H^{\sigma\nu}
\end{equation}
contains the Poynting vector and the direction of the wave vector,
\begin{align}
\label{poyntingdef}
 S^i &= M^{0i} = \epsilon^{ijk} E_j \mathcal H_k\ , \\ n^i &= M^{i0} = \gamma^{ij} \epsilon_{jkl} \mathcal D^k B^l \parallel k^i\ .
\end{align}
We recall that the original fields obey the constitutive relations of vacuous space and thus trivially $M^{0i} = M^{i0}$. From the transformation rules \eqref{trtildedef} the transformed tensor is found as
\begin{equation}
\begin{split}
 \tilde M^{\mu\nu} &= - \bar s \frac{\sqrt{-\bbar g}}{\sqrt{-g}} g^{\mu\rho} \bar F_{\rho\sigma} \bbar H^{\sigma\nu}\\ &= - \bar s \frac{\sqrt{-\bbar g}}{\sqrt{-g}} g^{\mu\rho} \frac{\partial x^\lambda}{\partial \bar x^\rho} F_{\lambda \tau} \frac{\partial x^\tau}{\partial \bar x^\sigma}  \frac{\partial \bbar x^\sigma}{\partial x^\alpha} H^{\alpha \beta} \frac{\partial \bbar x^\nu}{\partial x^\beta}\ .
 \end{split}
\end{equation}
The transformation law of $M^{\mu\nu}$ encodes in a geometric language how energy flux and phase velocity behave in a medium. For simplicity let us now concentrate on media without bi-anisotropic couplings, in other words we allow for general spatial transformations as well as stretchings and reversal of time, but keep $\bar g_{0i} = \bbar g_{0i} = 0$. Then we find for the transformed space vectors (cf.\ Eqs.\ \eqref{tildeE}--\eqref{tildeH}):
\begin{align}
\label{tildeS}
 \tilde S^i = \tilde M^{0i} &= - \bar s \bbar s \bbar \sigma \frac{\sqrt{-\bbar g_{00}}}{g_{00}} \frac{\partial x^0}{\partial \bar x^0} \epsilon^{ijk} \frac{\partial x^m}{\partial \bar x^j} E_m \frac{\partial x^n}{\partial \bbar{x}^k} \mathcal H_n \\
\label{tilden}
 \tilde n^i = \tilde M^{i0} &= \bar \sigma \frac{\sqrt{-\bbar g}\sqrt{\bar \gamma}}{\sqrt{\gamma}} \frac{\partial \bbar x^0}{\partial x^0} \gamma^{ij} \epsilon_{jkl} \frac{\partial \bbar x^k}{\partial x^m} \mathcal D^m \frac{\partial \bar x^l}{\partial x^n} B^n
\end{align}
The transformation of the Poynting vector may be abbreviated as
\begin{equation}
 \tilde S^i = T^{ijk} E_j \mathcal H_k\ ,
\end{equation}
and it is then easily seen that $n_i$ transforms as
\begin{equation}
 \tilde n_i = \bar \sigma \bar s \bbar \sigma \bbar s \frac{\sqrt{\bbar \gamma \bar \gamma}}{2 \gamma} g_{00} \frac{\partial \bbar x^0}{\partial x^0} \frac{\partial \bar x^0}{\partial x^0} U_{ikj} D^j B^k\ ,
\end{equation}
where $U_{ijk}$ is the inverse of $T^{ijk}$ in the sense of
\begin{equation}
 U_{ijk} T^{ljk} = \delta_i^l\ .
\end{equation}
While these formulae might look cumbersome, their geometric interpretation actually is quite straightforward. In the case of diffeomorphism transforming materials, $\bar g_{\mu\nu} = \bbar g_{\mu\nu}$, Eq.\ \eqref{tildeS} states that $S^i$ behaves under purely spatial transformations as a covector \cite{Pendry:2006Sc}, while $n_i$ from Eq.\ \eqref{tilden} behaves as a vector:
\begin{align}
 \tilde S^i &= \bar s \frac{\sqrt{-\bar g}}{\sqrt{-g_{00}}\sqrt{-g}} \frac{\partial x^0}{\partial \bar x^0} \frac{\partial \bar x^i}{\partial x^j} S^j\ , \\
 \tilde n_i &= \bar s \sqrt{-\bar g} \frac{\partial \bar x^0}{\partial x^0} \frac{\partial x^j}{\partial \bar x^i} n_j\ .
\end{align}
Of course, the relative orientation of $S^i$ and $n^i$ is preserved under the diffeomorphisms, but this is no longer true for $\tilde S^i$ and $\tilde n^i$, since indices are raised/lowered by the space metric $\gamma_{ij}$ in laboratory space as opposed to $\bar \gamma_{ij}$ in electromagnetic space.

For triple spacetime metamaterials no linear transformation $\tilde S^i = T^{i}{}_j S^j$ exists. This makes the interpretation a little bit more complicated, but at the same time is the source of the numerous additional possibilities within this generalized setup. In general, the value of the element $T^{ijk}$ defines the component of the Poynting vector in direction $x^i$ as generated by electric and magnetic fields that point in the original space in the directions $x^j$ and $x^k$, resp. In this way it is easy to engineer the direction of the Poynting vector in the medium for a given polarization of the incoming wave in vacuum. Similar conclusions apply for the transformation matrix $U_{ijk}$, with the notable restriction that $\tilde n^i$ can be parallel or anti-parallel to $k^i$. Whether $(\tilde{\mathcal D}^i,\tilde B^j,\tilde k^l)$ form a right- or left-handed triple can be deduced from
\begin{equation}
 \epsilon_{ijk} \tilde{\mathcal D}^j \tilde B^k = \frac{\tilde k_i}{\tilde \omega} \tilde E_j \epsilon^{jk} \tilde{E}_k = - \bar s \frac{\sqrt{-\bbar g}}{\sqrt{\gamma} g_{\bar 0 \bbar 0}} \frac{\tilde k_i}{\tilde \omega} \tilde E_j \gamma^{\bbar j \bar k} \tilde{E}_k\ .
\end{equation}

\subsection{Wave vector and dispersion relations}
\label{sec:wavevector}
While the above relations correctly reproduce the direction of the Poynting and the wave vector, they cannot distinguish between propagating and evanescent modes. Consider as an example the following transformation:
\begin{align}
 \bar x^0 &= x^0\ , & \bar x^i &= x^i\ , & \bbar x^0 &= -x^0\ , & \bbar x^i &= x^i\ .
\end{align}
From Eqs.\ \eqref{epstilde2} and \eqref{mutilde2} it is found that this is a homogeneous material with $\epsilon = -1$ and $\mu = 1$. As fields in vacuo, $\vec E = \vec{\mathcal D}$ and $\vec B = \vec{\mathcal H}$, we consider a monochromatic wave
\begin{align}
\label{mono1}
 \vec E &= \vec e e^{i(\vec k \vec x - \omega t)} + \mbox{c.c.}\ , \qquad \vec k \cdot \vec e = 0\ , \\
 \label{mono2}
 \vec B &= \vec b e^{i(\vec k \vec x - \omega t)} + \mbox{c.c.}\ , \qquad \vec b = \frac{1}{\omega} \vec k \times \vec e\ .
\end{align}
After the transformation the fields $\vec{\Tilde E}$, $\vec{\Tilde B}$ and $\vec{\Tilde{\mathcal D}}$, $\vec{\Tilde{\mathcal H}}$ refer to the original fields at different time instances:
\begin{align}
\label{ex1}
 \vec{\Tilde E}\left(\tilde x^\mu=\bar x^\mu(x)\right) &= \vec E(\vec x,t)\ , \\ \vec{\Tilde B}\left(\tilde x^\mu = \bar x^\mu(x)\right) &= \vec B(\vec x,t)\ , \\
 \vec{\Tilde{\mathcal D}}\left(\tilde x^\mu = \bbar x^\mu(x)\right) &= -\vec{\mathcal D}(\vec x, -t)\ , \\ \label{ex2} \vec{\Tilde{\mathcal H}}\left(\tilde x^\mu = \bbar x^\mu(x)\right) &= - \vec{\mathcal H}(\vec x, -t)\ .
\end{align}
Of course, Maxwell's equations are satisfied by the new fields by construction. Still, the partial exchange of positive and negative angular frequencies has important implications in the dispersion relation as any propagating wave in vacuo becomes evanescent in the medium and vice versa.

Though this behavior may not appear immediate when transforming the monochromatic wave \eqref{mono1} and \eqref{mono2} with \eqref{ex1}--\eqref{ex2}, it can be made explicit from geometric quantities as well. Indeed, from the relativistic wave equation \cite{Post}
\begin{equation}
 D_\nu \chi^{\mu\nu\rho\sigma} D_\rho A_\sigma = - J^\nu
\end{equation}
it follows straightforwardly that ``triple-space metamaterials'' in the absence of charges and currents and in the limit of approximate homogeneity obey the dispersion relation
\begin{equation}
 g^{\bbar \mu \bar \nu} k_\mu k_\nu = 0\ , \qquad k_\mu = (\omega, \vec{k})\ .
\end{equation}
In our example the partial reversal of time yields $g^{\bbar 0 \bar 0} = 1$ and thus $\omega^2 + \vec k^2 = 0$.

\section{Non-invariant transformations}
\label{sec:fieldtransforming}
Within the approaches to transformation media discussed so far invariant transformations of the equations of motions were used exclusively. This means that the transformations ``do not introduce charges or currents'', in other words the transformation medium based on a source free vacuum solution will be source free as well. What happens if this restriction is abandoned? Still insisting on a constitutive relation of the form \eqref{linearH} this suggests the following definition:
\begin{definition}
\label{def:general}
 A transformation medium is defined by an arbitrary linear transformation applied to a (not necessarily source free) vacuum solution of Maxwell's equations. The linear transformation constitutes the media properties as well as charges and currents of the transformation medium.
\end{definition}
Though not in its most general form, this approach was proposed in \cite{Tretyakov:2007Me,Tretyakov:2008Gf}. Starting from the vacuum relation \eqref{chivac} the most general linear relation can be achieved by the field transformations \footnote{Alternatively, one could start from a relation of the form
\[
 \bar H^{\mu\nu} = A^{\mu\nu}{}_{\rho\sigma} H^{\rho\sigma} + B^{\mu\nu\rho\sigma} F_{\rho\sigma}
\]
but the form \eqref{generalFT} appears more transparent to us.}
\begin{equation}
\label{generalFT}
 \bar H^{\mu \nu} = \Omega^{\mu \nu}{}_{\rho \sigma} H^{\rho \sigma}\ , \qquad F_{\mu \nu} = \Psi_{\mu \nu}{}^{\rho \sigma} \bar F_{\rho \sigma}\ ,
\end{equation}
with the transformed $\bar \chi$,
\begin{equation}
 \bar \chi^{\mu\nu\rho\sigma} = (\Omega \chi \Psi)^{\mu\nu\rho\sigma}\ .
\end{equation}
The original fields $F_{\mu\nu}$ and $H^{\mu\nu}$ by assumption are solutions to the equations of motion. If we allow besides the standard electric four-current $J^\mu$ also a magnetic four-current $J^\mu_M$, any transformation of the type \eqref{generalFT} can be mapped on a solution of the new equations ($\hat \Psi_{\mu \nu}{}^{\rho \sigma}$ is the inverse matrix $\hat \Psi_{\mu \nu}{}^{ \lambda\tau} \Psi_{\lambda\tau}{}^{\rho\sigma} = (\delta_\mu^\rho \delta_\nu^\sigma - \delta_\mu^\sigma \delta_\nu^\rho)/2$)
\begin{gather}
\label{eomtransf1}
 D_\mu \bar H^{\mu\nu} =  \bar J^\nu = D_\mu(\Omega^{\mu \nu}{}_{\rho \sigma} H^{\rho \sigma})\ , \\
\label{eomtransf2}
 \epsilon^{\mu\nu\rho\sigma} \partial_\nu \bar F_{\rho \sigma} = \bar J^\mu_M = \epsilon^{\mu\nu\rho\sigma} \partial_\nu (\hat \Psi_{\rho \sigma}{}^{\tau \lambda} F_{\tau \lambda})\ ,
\end{gather}
provided appropriate currents are introduced. These transformations in general are not symmetry transformations and accordingly a source free solution is no longer mapped on another source free solution. The transformations \eqref{generalFT} and the ensuing equations of motion \eqref{eomtransf1} and \eqref{eomtransf2} are the relativistic form of the transformations proposed in Refs.\ \cite{Tretyakov:2007Me,Tretyakov:2008Gf}.

We recover the invariant transformations of Section \eqref{sec:triplespace} by introducing the restrictions
\begin{equation}
\label{diffrestriction}
 \Omega^{\mu \nu}{}_{\rho \sigma} = S^{\mu}{}_{\rho}S^{\nu}{}_{\sigma}\ , \qquad \Psi_{\mu \nu}{}^{\rho \sigma} = T_{\mu}{}^{\rho} T_{\nu}{}^{\sigma}\ .
\end{equation}
Let us first count the degrees of freedom in the transformations. $\chi^{\mu\nu\rho\sigma}$ is a rank four tensor, anti-symmetric in $(\mu,\nu)$ and $(\rho,\sigma)$ and thus has 36 independent components (20 components of the principal part, 15 of the skewon part and one axion coupling). The same applies to the transformation matrices $\Omega$ and $\Psi$. Restriction to diffeomorphisms according to Eq.\ \eqref{diffrestriction} reduces this number to 6 parameters for each, $\Omega$ and $\Psi$; the electric-magnetic rotation of Section \ref{sec:emduality} adds another parameter in form of a rotation angle. Here, another important difference between a transformation material according to Definition \ref{def:two} and \ref{def:general} emerges. In both cases the transformation yielding certain media properties is not unique. In the former case, however, different transformations are physically equivalent as they are connected by symmetry transformations (isometries of the laboratory metric $g_{\mu\nu}$). In the more general case of Eq.\ \eqref{generalFT} the different transformations need not be physically equivalent. As is immediate from Eq.\ \eqref{generalFT} a certain medium exhibiting sources due to non-invariant transformations can be designed using electric charges and currents, magnetic charges and currents or both, which clearly characterizes physically different situations with the same media properties $\chi$.

Does there exist the possibility of a geometric interpretation of Eq.\ \eqref{generalFT}? If this shall be possible space must be transformed differently for different components of $F^{\mu\nu}$ and $H_{\mu\nu}$. In fact, the most general linear transformation can be interpreted as a separate transformation of spacetime for each component of the two tensors. Let us provide a simplified example, where independent spatial transformations are applied to $\vec E$, $\vec B$, $\vec{\mathcal D}$ and $\vec{\mathcal H}$. Laboratory space is denoted by $x^i$, the electromagnetic spaces by $x_E^i$, $x_B^i$, $x_D^i$ and $x_H^i$, resp. Under time-independent spatial transformations all four fields transform as (co-)vectors and thus Eqs.\ \eqref{tildeE}--\eqref{tildeH} suggest the interpretations
\begin{align}
\label{nitrans1}
 \tilde{E}_i &= s_E \frac{\partial x^j}{\partial x_E^i} E_j\ , & \tilde B^i &= \frac{\sqrt{\gamma_B}}{\sqrt{\gamma}} \frac{\partial x_B^i}{\partial x^j} B^j\ ,  \\
\label{nitrans2}
 \tilde{\mathcal D}^i &= \frac{\sqrt{\gamma_D}}{\sqrt{\gamma}} \frac{\partial x_D^i}{\partial x^j} \mathcal D^j\ , & \tilde{\mathcal H}_i &= s_H \frac{\partial x^j}{\partial x_H^i} \mathcal H_j\ ,
\end{align}
yielding for permittivity and permeability
\begin{align}
 \epsilon^{ij} &= s_E \frac{\sqrt{\gamma_D}}{\sqrt{\gamma}} \frac{\partial x^i_D}{\partial x^k} \gamma^{kl} \frac{\partial x^j_E}{\partial x^l}\ , \\ \mu^{ij} &= s_H \frac{\sqrt{\gamma_B}}{\sqrt{\gamma}} \frac{\partial x^i_B}{\partial x^k} \gamma^{kl} \frac{\partial x^j_H}{\partial x^l}\ .
\end{align}

As is seen from \eqref{maxwell} and \eqref{maxwell2} Gauss' law for $\tilde B^i$ and $\tilde{\mathcal D}^i$ remain unchanged, while Faraday's and Amp\`ere's laws are changed according to
\begin{gather}
 \frac{\sqrt{\gamma_E}}{\sqrt{\gamma_B}} \frac{\partial x^i_E}{\partial x^j_B} \nabla_0 \tilde B^j + \epsilon^{ijk} \partial_j \tilde E_k = 0\ , \\
 \epsilon^{ijk} \partial_j \tilde{\mathcal H}_k - \frac{\sqrt{\gamma_H}}{\sqrt{\gamma_D}} \frac{\partial x^i_H}{\partial x^j_D} \nabla_0 \tilde{\mathcal D}^j = 0\ , 
\end{gather}
making the electric and magnetic currents
\begin{align}
 j^i &=  - \left(\delta^i_j - \frac{\sqrt{\gamma_H}}{\sqrt{\gamma_D}} \frac{\partial x^i_H}{\partial x^j_D}\right) \nabla_0 \tilde{\mathcal D}^j\ , \\ j_M^i &= \left(\delta^i_j - \frac{\sqrt{\gamma_E}}{\sqrt{\gamma_B}} \frac{\partial x^i_E}{\partial x^j_B}\right) \nabla_0 \tilde{B}^j
\end{align}
necessary. For completeness it should be mentioned that the transformations \eqref{nitrans1} and \eqref{nitrans2} allow a straightforward interpretation since each of the four Maxwell's equations still can be transformed as a whole. Taking even more general transformations, e.g.\ transforming each component of the electric and magnetic fields separately, does no longer allow this manipulation in a simple way and thus will make the derivation of the necessary media parameters more complicated.

\section{Conclusions}
In this paper we have introduced a generalization of the concept of diffeomorphism transforming media, the basis of transformation optics \cite{Pendry:2006Sc,Leonhardt:2006Sc,Leonhardt:2006Nj}. As basic idea we have found that spacetime can be transformed differently for the field strength tensor (containing $\vec E$ and $\vec B$) and the excitation tensor (encompassing $\vec{\mathcal D}$ and $\vec{\mathcal H}$). This extension allows design of non-reciprocal media, in particular permittivity and permeability need not longer be symmetric. Furthermore, this approach permits a geometric interpretation of indefinite media \cite{Smith:2003Pr,Smith:2004Im}.

Diffeomorphism transforming media are motivated by the wish to produce a medium that looks like a transformed but empty space. The basis of this interpretation is Fermat's principle applied to these media \cite{Leonhardt:2008Oe}: indeed it is found that in a transformation medium the light rays travel along trajectories as if the medium was a transformed, empty space. Still, the transformation medium in general is quite different from transformed empty space, if the conservation laws from the stress-energy-momentum tensor are considered. This aspect is even more important within the extension proposed here, as there exist two different transformed (electromagnetic) spaces and light rays don't follow the geodesics of any of them. We have shown that one can make a virtue out of necessity: the geometric approach does not just provide a tool to design the path of light in a medium, but equally well it may be used to design the behavior of (parts of) the stress-energy-momentum tensor, e.g.\ the direction of the Poynting vector, and/or the behavior of the wave vector. Here the proposed generalization offers many more possibilities compared to the known diffeomorphism transforming media. In particular we have derived the geometric relations that describe the transformation of the Poynting vector, of the direction of the wave vector as well as the dispersion relation.

Finally we have commented on a different route to generalize the notion of transformation media \cite{Tretyakov:2007Me,Tretyakov:2008Gf}. These field-transforming media are not based on invariant transformations of the equations of motion and consequently source free solutions of the original configuration are not mapped onto source free solutions of the new medium. We have shown that also this approach may be covered by a generalized concept of coordinate transformations. Still, there remains a fundamental difference between the approach of Refs.\ \cite{Tretyakov:2007Me,Tretyakov:2008Gf} and the one discussed here: While in the former case the transformations are ultra-local (the transformed fields at the point $x^\mu$ are defined in terms of the original fields at this point), in the latter they are essentially non-local, as the transformed fields at $\tilde x^\mu$ are related to the original fields at some $x^\mu \neq \tilde x^\mu$. The preferable approach depends on the specific problem at hand, also a combination of the two is conceivable.

\begin{acknowledgments}
 The author wishes to thank J.~Llorens Montolio for helpful discussions. This work profited a lot from fruitful discussion with C.~Simovski, S.A.~Tretyakov, I.S.~Nevedov, P.~Alitalo, M.~Qiu and M.~Yan during a cooperation of the Advanced Concepts Team of the European Space Agency with the Helsinki University of Technology and the Royal Institute of Technology (KTH). The cooperation was funded under the Ariadna program of ESA.
\end{acknowledgments}

\appendix*
\section{Covariant formulation}
\label{sec:conventions}
In this Appendix we present our notations and conventions regarding the covariant formulation of Maxwell's equations on a possibly curved manifold. For a detailed introduction to the topic we refer to the relevant literature, e.g.\ \cite{Landau2,Post}. Throughout the whole paper natural units with $\epsilon_0 = \mu_0 = c = 1$ are used.

Greek indices $\mu, \nu, \rho, \ldots$ are spacetime indices and run from 0 to 3, Latin indices $i,j,k,\ldots$ space indices with values from 1 to 3. For the metric we use the ``mostly plus'' convention, so the standard flat metric is $g_{\mu\nu} = \mbox{diag}(-1,1,1,1)$. If we interpret $x^0 = t$ with (laboratory) time the space metric can be obtained as \cite{Landau2}
\begin{equation}
 \gamma^{ij} = g^{ij}\ , \qquad \gamma_{ij} = g_{lk} - \frac{g_{0i}g_{0j}}{g_{00}}\ , \qquad \gamma^{ij} \gamma_{jk} = \delta^i_k\ .
\end{equation}
This implies as relation between the determinant of the spacetime metric, $g$, and the one of the space metric, $\gamma$,
\begin{equation}
\label{detrel}
 -g = -g_{00} \gamma
\end{equation}
The four dimensional Levi-Civita tensor is defined as
\begin{align}
 \epsilon_{\mu\nu\rho\sigma} &= \sqrt{-g}[\mu\nu\rho\sigma]\ , & \epsilon^{\mu\nu\rho\sigma} &= - \frac1{\sqrt{-g}}[\mu\nu\rho\sigma]\ ,
\end{align}
with $[0123] = 1$. Therefore the reduction of the four dimensional to the three dimensional tensor reads
\begin{equation}
 \epsilon_{0ijk} = \sqrt{-g_{00}} \epsilon_{ijk}\ , \qquad \epsilon^{0ijk} = - \frac1{\sqrt{-g_{00}}} \epsilon^{ijk}\ .
\end{equation}
An additional complication arises in the definition of $\bar \epsilon_{ijk}$ and $\bbar \epsilon_{ijk}$, since the orientation of the spacetime manifold may change without changing the orientation of space (e.g.\ by a mapping $\bar t = -t$.) Therefore the corresponding relations should be written as
\begin{align}
 \bar \epsilon_{0ijk} &= \bar \sigma \sqrt{-\bar g_{00}} \bar \epsilon_{ijk}\ , & \bbar \epsilon_{0ijk} &= \bbar \sigma \sqrt{-\bbar g_{00}} \bbar \epsilon_{ijk}\ ,
\end{align}
where $\bar \sigma = +1$ if space and spacetime have the same orientation and $\bar \sigma = -1$ otherwise.

The field strength tensor $F_{\mu\nu}$ encompasses the electric field and the magnetic induction, the excitation tensor $H^{\mu\nu}$ the displacement vector and the magnetic field with the identification:
\begin{align}
\label{spacevec1}
 E_i &= F_{0i}\ , & B^i &= - \frac{1}{2} \epsilon^{ijk} F_{jk}\ , \\
\label{spacevec2}
 \mathcal D^{i} &= - \sqrt{-g_{00}} H^{0i}\ , & \mathcal H_i &= -\frac{\sqrt{-g_{00}}}{2} \epsilon_{ijk} H^{jk}\ .
\end{align}
Finally, electric charge and current are combined into a four-current $J^{\mu} = (\rho/\sqrt{-g_{00}}, j^i/\sqrt{-g_{00}})$. $F_{\mu\nu}$ and $H^{\mu\nu}$ are tensors, thus under the transformations of Section \ref{sec:triplespace} they behave as
\begin{align}
 \bar F_{\mu\nu} &= \frac{\partial x^\rho}{\partial \bar x^{\mu}} F_{\rho\sigma} \frac{\partial x^\sigma}{\partial \bar x^{\nu}}\ , & \bbar{H}^{\mu\nu} &= \frac{\partial \bbar x^{\mu}}{\partial x^{\rho}} H^{\rho\sigma} \frac{\partial \bbar x^{\nu}}{\partial x^{\sigma}}\ .
\end{align}
This implies for the transformed space vectors in laboratory space
\begin{align}
\label{tildeE}
\begin{split}
 \tilde{E}_i &= \bar s \biggl(\left(\frac{\partial x^0}{\partial \bar x^0} \frac{\partial x^j}{\partial \bar x^i} - \frac{\partial x^0}{\partial \bar x^i} \frac{\partial x^j}{\partial \bar x^0} \right) E_j\\ &\phantom{= \bar s \biggl(} - \frac{\partial x^j}{\partial \bar x^0} \frac{\partial x^k}{\partial \bar x^i} \epsilon_{jkl} B^{l}\biggr)\ , \end{split} \\
 \label{tildeB}
 \tilde B^i &= \bar \sigma \frac{\sqrt{\bar \gamma}}{\sqrt{\gamma}} \frac{\partial \bar x^i}{\partial x^j} B^j - \bar s \epsilon^{ijk} \frac{\partial x^0}{\partial \bar x^j} \frac{\partial x^l}{\partial \bar x^k} E_l\ , \\
 \label{tildeD}
 \begin{split}
 \tilde{\mathcal D}^i &= \frac{\sqrt{-\bbar{g}}}{\sqrt{-g}} \biggl(\left(\frac{\partial \bbar x^0}{\partial x^0} \frac{\partial \bbar x^i}{\partial x^j} - \frac{\partial \bbar x^0}{\partial  x^j} \frac{\partial \bbar x^i}{\partial x^0} \right) \mathcal D^j\\ &\phantom{= \frac{\sqrt{-\bbar{g}}}{\sqrt{-g}} \biggl(} + \frac{\partial \bbar x^0}{\partial x^j} \frac{\partial \bbar x^i}{\partial x^k} \epsilon^{jkl} \mathcal H_{l}\biggr)\ ,
 \end{split} \\
 \label{tildeH}
 \tilde{\mathcal H}_i &= \bbar s \bbar \sigma \frac{\sqrt{-\bbar{g}_{00}}}{\sqrt{-g_{00}}}\frac{\partial  x^j}{\partial \bbar x^i} \mathcal H_j + \frac{\sqrt{-\bbar{g}}}{\sqrt{-g}} \epsilon_{ijk} \frac{\partial \bbar x^j}{\partial x^0} \frac{\partial \bbar x^k}{\partial  x^l} \mathcal D^l\ . 
\end{align}

We characterize the general linear, lossless media usually by means of the Tellegen relations
\begin{align}
\label{tellegen}
 \mathcal D^i &= \epsilon^{ij} E_j + \kappa^{ij} \mathcal H_j\ , & B^i &= \mu^{ij} \mathcal H_j + \xi^{ij} E_j\ .
\end{align}
In terms of field strength and excitation tensor the media relations become the Boys-Post relation 
\begin{equation}
\label{linearHapp}
 H^{\mu \nu} = \frac{1}{2} \chi^{\mu\nu\rho\sigma} F_{\rho\sigma}\ ,
\end{equation}
where $\chi$ must be invertible with inverse
\begin{equation}
  \hat \chi_{\mu \nu \lambda\tau} \chi^{\lambda\tau\rho\sigma} = (\delta_\mu^\rho \delta_\nu^\sigma - \delta_\mu^\sigma \delta_\nu^\rho)\ .
\end{equation}
By virtue of Eqs.\ \eqref{spacevec1} and \eqref{spacevec2} Eq.\ \eqref{linearHapp} may be written as
\begin{equation}
 \begin{pmatrix}
 \mathcal H_i \\ \mathcal D^i
\end{pmatrix}
=
\begin{pmatrix}
 \mathfrak C_i{}^j & \mathfrak B_{ij}\\ \mathfrak A^{ij} & \mathfrak D^{i}{}_{j}
\end{pmatrix}
\begin{pmatrix}
 E_j \\ B^j
\end{pmatrix}\ ,
\end{equation}
with
\begin{align}
\label{ABdef}
 \mathfrak A^{ij} &= - \sqrt{-g_{00}} \chi^{0i0j}\ , \\ \mathfrak B_{ij} &= \frac{1}{8} \sqrt{-g_{00}} \epsilon_{ikl} \epsilon_{jmn} \chi^{klmn}\ ,\\
 \mathfrak C_i{}^j &= - \frac{1}{2} \sqrt{-g_{00}} \epsilon_{ikl}\chi^{kl0j}\ , \\ \label{CDdef} \mathfrak D^{i}{}_{j} &= \frac{1}{2} \sqrt{-g_{00}} \epsilon_{jkl} \chi^{0ikl}\ .
\end{align}
The Tellegen and Boys-Post formulations are related by
\begin{align}
 \epsilon^{ij} &= \left(\mathfrak A - \mathfrak D \mathfrak B^{-1} \mathfrak C\right)^{ij} & \kappa^{ij} &= \left(\mathfrak D \mathfrak B^{-1}\right)^{ij} \\
 \mu^{ij} &= (\mathfrak B^{-1})^{ij} & \xi^{ij} &= - \left(\mathfrak B^{-1} \mathfrak C\right)^{ij}
\end{align}
Finally, we mention that the rank 4 tensor $\chi^{\mu\nu\rho\sigma}$ may be decomposed as \cite{Hehl:2003,Hehl:2005hu}
\begin{equation}
\label{chidecomp}
 \chi^{\mu\nu\rho\sigma} = {}^{(1)}\chi^{\mu\nu\rho\sigma} + \epsilon^{\mu\nu\lambda[\rho} S_{\lambda}{}^{\sigma]} - \epsilon^{\rho\sigma\lambda[\mu} S_{\lambda}{}^{\nu]} + \alpha \epsilon^{\mu\nu\rho\sigma}\ ,
\end{equation}
where the principal part ${}^{(1)}\chi^{\mu\nu\rho\sigma}$ has no part completely anti-symmetric in its indices and is symmetric under the exchange $(\mu,\nu)\leftrightarrow(\rho,\sigma)$. The principal part has been discussed extensively in \cite{Post}. $S_\mu{}^\nu$ was introduced in Refs.~\cite{Hehl:2003,Hehl:2005hu} as skewon part (related to chiral properties of the material \cite{Hehl:2005hu,lindell}), while $\alpha$ represents the well-known axion coupling \cite{Wilczek:1987mv}.

\bibliography{bibtsp}

\end{document}